\documentclass[sigconf]{acmart}

\usepackage{booktabs} 

\setcopyright{acmlicensed}

\usepackage{url}

\usepackage{color}

\usepackage{rotating}
\usepackage{tablefootnote}
\usepackage[bottom]{footmisc}

\usepackage{multirow}
\usepackage{makecell}
\usepackage{tabularx}
\usepackage{threeparttable}

\usepackage{graphicx} 
\usepackage{stfloats}
\usepackage{float}
\usepackage{epstopdf}
\usepackage{epsfig}
\usepackage{adjustbox}
\usepackage{tabularx}

\usepackage{balance}

\usepackage{tikz}
\usepackage{pgfplots, pgfplotstable}
\pgfplotsset{compat=1.12}
\usepackage{filecontents}
\usetikzlibrary{shapes,arrows,positioning}

\usepackage{xspace}

\let\labelindent\relax
 \usepackage{enumitem}

\usepackage{verbatim}
\usepackage{csquotes}
\usepackage{fmtcount}

\newcommand{\paragraphX}[1]{\vskip 4pt \noindent \textbf{#1} \hskip .05in}

\newcommand\hmm[1]{\ifnum\ifhmode\spacefactor\else2000\fi>1005 \uppercase{#1}\else#1\fi}
\newcommand{\wf}{\hmm{w}ebsite fingerprinting\xspace}
\newcommand{\hs}{\hmm{o}nion service\xspace}
\newcommand{\hses}{\hmm{o}nion services\xspace}
\newcommand{\tikzcircle}[2][red,fill=red]{\tikz[baseline=-0.5ex]\draw[#1,radius=#2] (0,0) circle ;}%
\newcommand{\reddot}{\tikzcircle[fill=gray]{2pt}}

\newcommand{\bo}[1]{\textcolor{red}{}} \newcommand{\mj}[1]{\textcolor{green}{}} \newcommand{\ga}[1]{\textcolor{blue}{}} \newcommand{\rg}[1]{\textcolor{yellow}{}} \newcommand{\cd}[1]{\textcolor{pink}{}}

\setcopyright{acmlicensed}

\acmDOI{10.1145/3133956.3134005}
\acmISBN{978-1-4503-4946-8/17/10}

\acmConference{CCS '17}{October 30-November 3, 2017}{Dallas, TX, USA}
\copyrightyear{2017} 
\acmYear{2017} 
\acmPrice{15.00}

\usepackage{hyperref}
\begin{document}
\title{How Unique is Your .onion? \\An Analysis of the Fingerprintability of Tor Onion Services}

\author{Rebekah Overdorf}
\affiliation{%
  \institution{Drexel University}
  \city{Philadelphia} 
  \state{Pennsylvania} 
}
\email{rebekah.overdorf@drexel.edu}

\author{Marc Juarez}
\affiliation{%
  \institution{ESAT-COSIC and imec KU Leuven}
  \city{Leuven} 
  \country{Belgium} 
}
\email{marc.juarez@kuleuven.be}

\author{Gunes Acar}
\affiliation{%
  \institution{imec-COSIC KU Leuven}
  \city{Leuven} 
  \country{Belgium} 
}
\email{gunes.acar@esat.kuleuven.be}

\author{Rachel Greenstadt}
\affiliation{
  \institution{Drexel University}
  \city{Philadelphia} 
  \state{Pennsylvania} 
}
\email{rachel.a.greenstadt@cs.drexel.edu}

\author{Claudia Diaz}
\affiliation{%
  \institution{imec-COSIC KU Leuven}
  \city{Leuven} 
  \country{Belgium} 
}
\email{claudia.diaz@esat.kuleuven.be}
\renewcommand{\shortauthors}{R. Overdorf et al.}

\begin{abstract}
Recent studies have shown that Tor onion (hidden) service websites are particularly vulnerable to \wf attacks due to their limited number and sensitive nature. In this work we present a multi-level feature analysis of onion site fingerprintability, considering three state-of-the-art \wf methods and 482 Tor \hses, making this the largest analysis of this kind completed on \hses to date.  

Prior studies typically report average performance results for a given \wf method or countermeasure.  We investigate which sites are more or less vulnerable to fingerprinting and which features make them so. We find that there is a high variability in the rate at which sites are classified (and misclassified) by these attacks, implying that average performance figures may not be informative of the risks that \wf attacks pose to particular sites.

We analyze the features exploited by the different \wf methods and discuss what makes \hs sites more or less easily identifiable, both in terms of their traffic traces as well as their webpage design. We study misclassifications to understand how \hses sites can be redesigned to be less vulnerable to \wf attacks. Our results also inform the design of \wf countermeasures and their evaluation considering disparate impact across sites.
\end{abstract}




\maketitle

\section{Introduction}~\label{sec:intro}






\wf attacks apply supervised classifiers to network traffic traces to identify
patterns that are unique to a web page. These attacks can circumvent the protection afforded by  encryption~\cite{Cheng1998,Hintz2003,Liberatore2006,Sun2002} and the metadata protection of anonymity systems such as Tor~\cite{Dingledine2004,Herrmann2009}. To carry out the attack the adversary first visits the websites, 
records the network traffic of his own visits, and extracts from it a  \emph{template} or
\emph{fingerprint} for each site. Later, when the victim user connects to the site (possibly through Tor), the adversary observes the victim's traffic and compares it to the previously recorded templates, trying to find a match. \wf can be deployed by adversaries with modest resources who have access to the communications between the user and the Tor entry guard. There are many entities in a position to access this communication, including wireless router owners, local network administrators or eavesdroppers, Internet Service Providers (ISPs), and Autonomous Systems (ASes), among other network intermediaries.


Despite the high success rates initially reported by \wf attacks~\cite{Cai2012,Wang2013}, their practicality in the real-world remains uncertain. A 2014 study showed that the success of the attacks is significantly lower in realistic scenarios than what is reported by evaluations done under artificial laboratory conditions~\cite{Juarez2014}. Moreover, using a very large world of websites, Panchenko et al. showed that \wf attacks do not scale to the size
of the Web~\cite{Panchenko2016}, meaning that, in practice, it is very hard  for an adversary to use this attack to recover the browsing history of a Tor user. 

Kwon et al.\ demonstrated, however, that a \wf adversary can reliably distinguish \hs connections from other Tor connections~\cite{Kwon2015}. This substantially reduces the number of sites to consider when only targeting \hses, as the universe of \hses is orders of magnitude smaller than the web,  which makes \wf attacks  potentially effective in practice. In addition, \hses are used to host sensitive content such as whistleblowing platforms and activist blogs, making  \wf attacks on this sites particularly attractive, and potentially very damaging~\cite{Cherubin2017}. For these reasons, we focus our analysis on \hses rather than the whole web. 

In this work we choose to model the set of \hses as a closed world. Our dataset contains as many landing pages of the hidden service world as was possible for us to collect at the time. After removing pages with errors and pages that are duplicates of other sites, we were left with a sanitized dataset of 482 out of the 1,363 \hses that were crawled. While the exact size of the complete \hs world cannot be known with certainty,~\texttt{onionscan} was  able to find 4,400 \hses on their latest scan (this number is not sanitized for faulty or duplicated sites)~\cite{Lewis2017}. This indicates that our set, while incomplete, contains a significant portion of the \hs world. We consider that an actual attacker can compile an exhaustive list of \hses, which would effectively yield a closed world scenario, since, once the adversary establishes that a user is visiting a \hs, the \hs in question will be one on the adversary's list. We note that closed world models are not realistic when considering the entire web, rather than just \hses. 




Prior evaluations of \wf attacks and defenses report aggregate metrics such as average classifier accuracy. However, we
find that some websites have significantly more distinctive fingerprints than others across classifiers, and that average metrics such as overall classifier accuracy cannot capture this diversity. 

In this work, we study what we call the \emph{fingerprintability} of websites and investigate what makes a page more vulnerable to \wf. This issue has practical relevance because adversaries interested in identifying visits to a particularly sensitive site may not care about the accuracy of the classifier for other sites, and thus the fingerprintability of that specific site matters. Similarly, the administrators of onion services likely care more about the vulnerability of {\emph {their}} users to fingerprinting attacks, rather than the average vulnerability of a \hses to the attack. We extract lessons from  our analysis to provide recommendations to \hs designers to better protect their sites against \wf attacks, including an analysis of a high profile SecureDrop instance.

The contributions of this study are:

{\bf Large .onion study. \footnote{This data along with the code used for analysis in this work is available at \url{https://cosic.esat.kuleuven.be/fingerprintability/}}} We collected the largest dataset of \hses for \wf to date and evaluated the performance of three state-of-the-art classifiers in successfully identifying \hs sites. For comparison, previous studies considered worlds of 30~\cite{Hayes2016}  or 50~\cite{Kwon2015,Cherubin2017} \hses, an order of magnitude smaller than our study, that analyses 482 \hses. 


{\bf Fingerprintability matters.} While the average accuracy achie\-ved by the classifiers is 80\%, we found that some sites are consistently misclassified by \emph{all} of the methods tested in this work, while others are consistently identified correctly, and yet others provide mixed results. In particular, 47\% of sites in our data set are classified with greater than 95\% accuracy, while 16\% of sites were classified with less than 50\% accuracy. Throughout this paper, we use the term \emph{fingerprintable} to mean how many of the visits are correctly classified. Depending on the requirements of the specific analysis, we use different ways to distinguish more and less fingerprintable sites. This includes comparing top 50 sites to bottom 50 sites or taking sites with $F1<0.33$ as less fingerprintable and sites with $F1>0.66$ as more fingerprintable.


\textbf{Errors made by different methods are correlated.}
Fully 31\% of misclassified instances were misclassified by all three classifiers. This implies that weaknesses of the individual classifiers cannot be fully overcome using ensemble methods. We nonetheless propose an \textit{ensemble} that combines all three classifiers, slightly improving the results offered by the best individual classifier. 

\textbf{Novel feature analysis method.} We present a method for analyzing fingerprintability that considers the relationship between the inter-class variance and intra-class variance of features across sites. The results of this analysis explain which features make a site fingerprintable, independently of the classifier used.

\textbf{Size matters.} We show that size-based features are the most important in identifying websites and that when sites are misclassified, they are typically confused with sites of comparable size. We show that large sites are consistently classified with high accuracy. 

\textbf{Dynamism matters for small sites.} While large sites are very fingerprintable, some small sites are harder than others to classify. We find that misclassified small sites tend to have more variance, and that features related to size variability are more distinguishing in sets of small sites. Put simply, smaller sites that change the most between visits are the hardest to identify.

\textbf{Analysis of site-level features.} Site-level features are website design features that cannot be (directly) observed in the encrypted stream of traffic but can be tweaked by the \hs operators. We identify which site-level features influence fingerprintability and we provide insights into how \hses can be made more robust against \wf attacks. 



{\bf Insights for Adversarial Learning.} \wf is a dynamic, adversarial learning problem in which the attacker aims to classify a traffic trace and the defender aims to camouflage it, by inducing misclassifications or poisoning the learning system. In the parlance of adversarial learning~\cite{Barreno2006}, we have conducted an exploratory attack against three different approaches, to help site owners and the Tor network design better causative attacks. A causative attack is an attack against a machine learning system that manipulates the training data of a classifier. Most adversarial learning approaches in the literature consider the adversary to be the evader of the learning system, not the learner. However, this is not the case in \wf nor in many other privacy problems. For this reason, most adversarial learning studies investigate an attack on a specific learning algorithm and feature set. In contrast, we study the three top-performing learners and introduce a classifier-independent feature analysis method to  study the learnability of a particular class (a web page).

\section{Background and Related Work}~\label{sec:related}


Encryption alone does not hide source and destination IP addresses, which can reveal the identities of the users and visited website. Anonymous communications systems such as Tor~\cite{Dingledine2004} route communications through multiple relays, concealing the destination server's address from network adversaries. Moreover, Tor supports \hses which can be reached through Tor while concealing the location and network address of the server. 


\textit{Website fingerprinting} is a traffic analysis attack that allows an attacker to recover the browsing history of a user from encrypted and anonymized streams.  Prior work has studied the effectiveness of this attack on HTTPS~\cite{Cheng1998}, encrypted web proxies~\cite{Sun2002,Hintz2003}, OpenSSH~\cite{Liberatore2006}, VPNs~\cite{Herrmann2009}, and various anonymity systems such as Tor and JAP~\cite{Herrmann2009}. We focus on Tor because it is, with more than two million daily users~\cite{tor-metrics}, the most popular anonymous communications system.

\begin{figure}[b]
 \centering
 \includegraphics[scale=1.05]{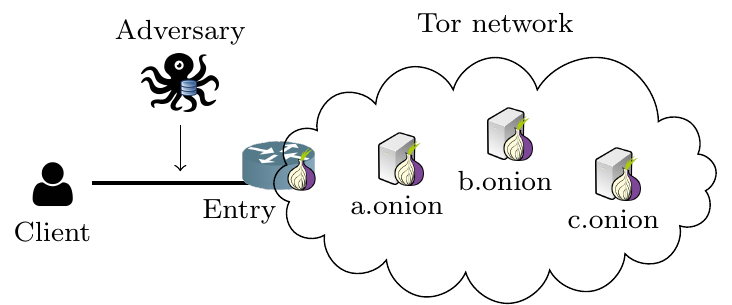}
 \caption{The client visits an \hs site over the Tor network. The adversary has
 access to the (encrypted) link between the client and the \emph{entry} to the
 Tor network. For clarity, we have omitted the six-hop circuit between the
 client and the \hs. The attacker cannot observe traffic beyond the
 entry node.}
  \label{fig:threat_model}
\end{figure}

In \wf the adversary is a network eavesdropper who can identify the user by her IP address, but who does not know which website the user is visiting (see \autoref{fig:threat_model}). The attacker cannot decrypt the communication, but can record the network packets generated by the activity of the user. To guess the web page that the user has downloaded, the attacker compares the traffic recorded from the user with that of his own visits to a set of websites. The best match is found using a statistical classifier.

\wf attacks are based on supervised classifiers where the
training instances are constructed from the traffic \emph{samples} or
\emph{traces} the adversary collects browsing sites of interest with with Tor, and the test samples are traces presumably captured from Tor users'\ traffic. 
Next, we will give an overview of \wf attacks that have been proposed in the literature.

\subsection{Attacks against Tor}

In 2009, Herrmann et al.\ proposed the first \wf attack against Tor, based on a Naive Bayes classifier and frequency distributions of packet lengths~\cite{Herrmann2009}. Their study only achieved an average accuracy of 3\% for 775 websites, but their attack was improved by Panchenko et al.\ who used a Support Vector Machine (SVM) and extracted additional features from traffic bursts to classify Herrmann et al.'s dataset with more than 50\% accuracy~\cite{Panchenko2011}.

Panchenko et al.'s study was also the first to perform an open-world evaluation of \wf attacks~\cite{Panchenko2011}. Prior work relied on a \emph{closed-world assumption}, which assumes that the universe of possible pages is small enough that adversary can train the classifier on all sites. The open-world evaluation is appropriate for a web environment as it accounts for users visiting pages that the classifier has not been trained on. Based on Herrman et al.'s dataset, Cai et al.~\cite{Cai2012} achieved more than 70\% accuracy in an open-world setting. Wang and Goldberg's~\cite{Wang2013} approach obtained over 90\% accuracy for 1,000 sites in an open world setting.

The results reported by these attacks were criticized for using experimental conditions that gave unrealistic advantages to the adversary, compared to real attack settings~\cite{Juarez2014}. However, new techniques have been shown to overcome some of those limitations, suggesting that attacks may be successful in the wild~\cite{Wang2016}.

Even though an open-world is a more realistic evaluation setting than a closed world for the web, our evaluation considers a closed world because: i) the universe of \hses is small enough that is feasible for an adversary to build a database of fingerprints for all existing \hses; and ii)  we
are interested in the best-case scenario for the adversary because we evaluate the vulnerability to \wf from a defender's point of view.

As in most prior work on \wf, we only consider the homepages of the websites and not inner pages within a website. We justify this for \hses by arguing that, given their unusable naming system and their shallowness in terms of not having a deep structure, it is reasonable to assume that visitors of \hses will land first on homepage more often than for regular sites before logging in or further interacting with the site.

In this paper, we focus only on \hses because a 2015 study showed that the \wf
adversary can distinguish between visits to \hses and regular
websites with high accuracy~\cite{Kwon2015}. Even though Panchenko et al.'s study shows that \wf does not scale to the Web, \wf has been identified as a potential threat for \hses for
two reasons~\cite{Cherubin2017}: first, in contrast to the Web's size, the \hs space's size may be sufficiently small for an adversary to build a fingerprint database for all existing \hses; second, onion services tend to host sensitive content and visitors of these sites may be subject to more serious, adverse consequences.


\subsection{State-of-the-art attacks}
We have selected three classifiers proposed in recent prior work for our study because they represent the most advanced and effective \wf attacks to date. Each attack uses different classification algorithms and feature sets, although they have some features in common. The details of each classifier are as follows:

\paragraphX{\bf Wang-kNN~\cite{Wang2014}:} Wang et al.\ proposed an attack
based on a k-Nearest Neighbors (k-NN) classifier that used more than 3,000
traffic features. Some of the most relevant features are the number of outgoing packets in spans of 30 packets, the lengths of the first 20 packets, and features that capture traffic \emph{bursts}, i.e., sequences of packets in the same direction.  They also proposed an algorithm to tune the weights of the custom distance metric used by the k-NN that minimizes the distance among
instances that belong to the same site. They achieved between 90\% to 95\% accuracy on a closed-world of 100 non-\hs websites~\cite{Wang2014}. Kwon et al.\ evaluated their own implementation of the attack for 50 \hs sites and obtained 97\% accuracy. 

\paragraphX{\bf CUMUL~\cite{Panchenko2016}:} Panchenko et al.\ designed CUMUL, an attack based on a Radial Basis Function kernel (RBF) SVM. Each feature instance is a 104-coordinate vector formed by the number of bytes and packets in each direction and 100 interpolation points of the cumulative sum of packet lengths (with direction). They report success rates that range between 90\% and 93\% for 100 regular sites. In addition, they collected the largest and most realistic dataset of non-\hs websites, including \emph{inner} pages of websites and popular links extracted from Twitter. They conclude
that \wf does not scale to such large dataset, as classification errors
increase with the size of the world.

\paragraphX{\bf k-Fingerprinting (k-FP)~\cite{Hayes2016}:} Hayes and Danezis's
k-FP attack is based on Random Forests (RF). Random Forests are ensembles
of decision trees that are randomized and averaged to reduce overfitting. In
the open-world, they use the leafs of the random forest to encode websites.
This allows them to represent websites in function of the outputs of the random forest, capturing the relative distance to pages that individual trees have confused with the input page. The instances extracted from the random forest are then fed into a k-NN classifier for the actual classification. The study uses a set of 175 features that includes variations of features in the literature as well as timing features such as the number of packets per second.
Hayes and Danezis evaluated the attack on a limited set of 30 \hses and obtained 90\% classification accuracy~\cite{Hayes2016}.

In the following subsection we provide an overview of prior results on features that has inspired the feature selection made by these three attacks. 

\subsection{Feature analysis for website fingerprinting}




We consider two types of features: \emph{network-level} and \emph{site-level} features. Network-level features are extracted from the stream of TCP packets and are the typical features used in \wf attacks. Site-level features are related to the web design of the site. These features are not available in the network traffic meta-data, but the adversary still has access to them by downloading the site.

Most \wf feature analyses have focused on network-level features and have evaluated their relevance for a specific classifier~\cite{Dyer2012,Cai2014b,Panchenko2011}. In particular, Hayes and Danezis~\cite{Hayes2016} perform an extensive feature analysis by compiling a comprehensive list of features from the \wf literature as well as designing new features. In order to evaluate the importance of a feature and rank it, they used the random forest classifier on which their attack is based.

Unlike prior work, our network-level feature analysis is classifier-independent, as we measure the statistical variance of features among instances of the same website (\emph{intra-class variance}) and among instances of different websites (\emph{inter-class variance}). 




\subsection{Website fingerprinting defenses}

Dyer et al.\ presented \emph{BuFLO}, a defense that delays real messages and adds dummy messages to make the traffic look constant-rate, thus concealing the features that \wf attacks exploit. They conclude that coarse-grained features such as page load duration and total size are expensive to hide with BuFLO and can still be used to distinguish websites ~\cite{Dyer2012}.

There have been attempts to improve BuFLO and optimize the padding at the end of the page download to hide the total size of the page~\cite{Cai2012,Cai2014a}. These defenses however incur high latency overheads that make them unsuitable for Tor. To avoid introducing delays, a \wf defense based solely on adding dummy messages was proposed by Juarez et al.~\cite{Juarez2016}. These defenses aim at crafting padding to obfuscate distinguishing features exploited by the attack. Instead, we look at sites and examine what makes them more or less fingerprintable. 

There are defenses specifically designed for Tor that operate at the application layer~\cite{Perry2011,Luo2011,Cherubin2017}. However, these defenses do not account either for feature analyses that can help optimize the defense strategy. Our study is the first to analyze the features at both the website and network layers. Based on our results, we discuss ways to reduce the fingerprintability of \hs sites and inform the design of server and client-side \wf defenses without requiring any changes to the Tor protocol itself.

\section{Data Collection and Processing}\label{sec:data_collection}

We used the \hs list offered by \texttt{ahmia.fi}, a search engine
that indexes \hses. We first downloaded a list of 1,363 \hs websites and found that only 790 of them were online using a shell
script based on \texttt{torsocks}. We crawled the homepage of the 790 online \hses.

Prior research on \wf collected traffic data by grouping
visits to pages into batches, visiting every page a number of times each
batch~\cite{Juarez2014,Wang2013}. All visits in a batch used the same Tor
instance but Tor was restarted and its profile wiped between batches, so that
visits from different batches would never use the same circuit. The  batches were used as cross-validation folds in the evaluation of the
classifier, as having instances collected under the same circuit in both
training and test sets gives an unfair advantage to the
attacker~\cite{Juarez2014,Wang2013}.

In this study, we used the same methodology to collect data, except that we
restarted Tor on every visit to avoid using the same circuit to download the same page multiple times. We ran the crawl on a cloud based Linux machine from a data center in the US in July 2016. The crawl took 14 days to complete which allowed us to take several snapshots of each \hs in time.


We used Tor Browser version 6.0.1 in combination with Selenium browser
automation library~\footnote{\url{http://docs.seleniumhq.org/}}. For each visit, we collected network
traffic, HTML source code of the landing page, and HTTP request-response
headers. We also saved a screenshot of each page.

We captured the network traffic traces using the \texttt{dumpcap}~\footnote{\url{https://www.wireshark.org/docs/man-pages/dumpcap.html}} command line tool. After each
visit, we filtered out packets that were not
destined to the Tor guard node IP addresses. Before each visit, we downloaded and processed the Tor network consensus with \texttt{Stem}~\footnote{\url{https://stem.torproject.org/}} to get the list of current guard IP addresses.


The HTML source code of the index page was retrieved using Selenium's
\texttt{page\_source} property. The source code and screenshots are used to extract site-level features, detect connection errors and duplicate sites. The HTTP requests and response headers are stored using a custom Firefox browser add-on. The add-on intercepted all HTTP requests, including the dynamically generated ones, using the~\texttt{nsIObserverService} of Firefox~\footnote{\url{https://developer.mozilla.org/en/docs/Observer_Notifications\#HTTP_requests}}.

Finally, we collected the logs generated by Tor Browser binary and Tor controller
 logs by redirecting Tor Browser's process output to a log file.

\subsection{Processing crawl data}
We ran several post-processing scripts to make sure the crawl data was useful
for analysis.


{\bf Remove offline sites.}
		Analyzing the collected crawl data, we removed 573 sites as they were
	found to be offline during the crawl. 

{\bf Remove failed visits.}
		We have also removed $14481$ visits that failed due to connection errors,
		possibly because some onion sites have intermittent uptime and are
		reachable temporarily. 

{\bf Outlier removal.} 
		We used Panchenko et al.'s outlier removal strategy to exclude packet captures of uncommon sizes compared to other visits to the
		same site~\cite{Panchenko2016}. This resulted in the removal of $5264$ visits.

{\bf Duplicate removal.}
		By comparing page title, screenshot and source code of different \hses, we found that some \hs websites are served on multiple
		\texttt{.onion} addresses. We eliminated $159$ duplicate sites by removing all copies of the site but one.

{\bf Threshold by instances per website.}
After removing outliers and errored visits, we had an unequal number of instances across different websites. Since the number of training instances can affect classifier accuracy, we set all websites to have the same number of instances. Most datasets in the literature have between 40 and 100 instances per website and several evaluations have shown that the accuracy saturates after 40 instances~\cite{Wang2013,Panchenko2016}. We set the threshold at 70 instances which is within the range of number of instances used in the prior work. Choosing a greater number of instances would dramatically decrease the final number of websites in the dataset. We removed 84 sites for not having a sufficient number of instances and removed 9,344 extra instances.

{\bf Feature Extraction.}
Following the data sanitization steps outlined above, we extract features used by the three classifiers. Further, we extract~\emph{site level} features using the HTML source, screenshot, HTTP requests and responses.
Site level features are explained in Section~\ref{sec:high-level}.

In the end, the dataset we used had 70 instances for 482 different onion sites.

\section{Analysis of Website Classification Errors}~\label{sec:misclassif}

This section presents an in-depth analysis of the successes and failures of the
three state-of-the-art \wf methods. This analysis helps identify which pages are the most fingerprintable and which are more likely to confuse the classifiers, giving insight into the nature of the errors produced by the classifiers.

\subsection{Classifier Accuracy}

Even though the classification problem is not binary, we binarize the problem by using a \emph{one-vs-rest} binary problem for each site: a True Positive (TP) is an instance that has been correctly
classified and False Positive (FP) and False Negative (FN) are both errors with respect to a fixed site $w$; a FP is an instance of another site that has been classified as $w$; a FN is an
instance of $w$ that has been classified as another site.

In the closed world we measure the accuracy using the F1-Score (F1). The
F1-Score is a \emph{complete} accuracy measure because it takes into account
both Recall (TPR) and Precision (PPV). More precisely, the F1-Score is the
harmonic mean of Precision and Recall: if either is zero, the F1-Score is zero as well,
and only when both achieve their maximum value, the F1-Score does so too. 

Note that there are the same \emph{total} number of FPs
and FNs, since a FP of $w_y$ that actually belongs to $w_x$ is at the same time
a FN of $w_x$. Thus, in the closed world the total F1-Score equals both Precision and Recall. However, when we focus on a particular site, the FP and
FN for that site are not necessarily the same (see \autoref{tab:top_sites}).

\begin{table}[htp]
 \centering
 \caption{Closed world classification results for our dataset of 482 \hses (33,740 instances in total).}
 \vspace{0.5cm}
 \label{tab:closed_world_scores}
  \begin{threeparttable}
    \resizebox{0.75\linewidth}{!}{%
   \begin{tabular} {lccccc} 
                      & k-NN     & CUMUL    & k-FP    \\
    \cmidrule{1-4} 
    \addlinespace
      TPR              & 69.97\%   & 80.73\%   & 77.71\% \\
    \addlinespace
      FPR              & 30.03\%   & 19.27\%    & 22.29\%   \\
    \addlinespace
   \end{tabular}}
  \end{threeparttable}
   \vspace{-0.3cm}
\end{table}

We have applied the classifiers to our dataset of 482 \hses and evaluated the
classification using 10-fold cross-validation. Cross-validation is a standard
statistical method to evaluate whether the classifier generalizes for instances
that it has not been trained on. In most cases, ten is the recommended
number of folds in the machine learning literature and the standard in prior
\wf work. The results for each classifier are summarized in
\autoref{tab:closed_world_scores} where we report the total number of TPs and
FPs and the average accuracy obtained in the 10-fold cross-validation. Thus, we
note that using TPR as an accuracy metric is sound in the closed world but, in
the open world, TPR is a partial measure of accuracy, as it does not take into
account Precision.

As we see in \autoref{tab:closed_world_scores}, while CUMUL and k-FP achieve
similar accuracies, the k-NN-based attack is the least accurate. Even though
these results are in line with other studies on \wf for
\hses~\cite{Cherubin2017}, we found some discrepancies with other evaluations
in the literature. For 50 sites, Hayes and Danezis obtain over 90\% accuracy
with k-FP~\cite{Hayes2016}, and Kwon et al.\ obtained 97\% with
k-NN~\cite{Kwon2015}. However, for the same number of sites and even more
instances per site, our evaluations of k-FP and k-NN only achieve 80\% maximum
accuracy. Since our results show that some sites are more fingerprintable than
others, we believe the particular choice of websites may account for this
difference: we randomly picked 50 sites from our set of 482 sites and even though Kwon et al.\, also used onion URLs from ahmia.fi, they do not explain how they picked the URLs for their evaluation.


\subsection{Classifier Variance}

In order to determine which features cause a site to be fingerprintable, we
look into two types of sites: i) sites that are easy to fingerprint,
i.e., sites that consistently cause the least amount of errors across all
classifiers; and ii) sites that are difficult to fingerprint, namely sites that
are most frequently misclassified across all three classifiers. In the
following sections, we compare the features of these two types of sites and
look for evidence that explains their different degree of fingerprintability. 

In our analysis, we evaluated the accuracy for
each website in isolation and ranked all the websites to find a threshold that divides them into the two types described above. We found that only 10 (in kNN) to 40 (in CUMUL) sites are perfectly classified, while the other sites have at least one misclassified instance -- some of them are consistently misclassified by all three classifiers. 

 
 
 

\begin{table}[tp]
 \centering
 \caption{The top five \hses by number of misclassification for each attack (repeating services in bold).}
 \vspace{0.5cm}
 \label{tab:top_sites}
  \begin{threeparttable}
  \resizebox{0.75\linewidth}{!}{%
   \begin{tabular}{cccccccc}
	   & URL (.onion)		& TP & FP  & FN     & F1  \\ 
    \hline
     \multirow{5}{*}{\begin{turn}{90}k-NN\end{turn}}
       & 4fouc$\ldots$  & 4  & 84  & 66  & 0.05 \\ 
       & \textbf{ykrxn}$\ldots$  & 3  & 62  & 67  & 0.04 \\ 
       & \textbf{wiki5k}$\ldots$  & 3  & 77  & 67  & 0.04 \\ 
       & ezxjj$\ldots$  & 2  & 76  & 68  & 0.03 \\ 
       & \textbf{newsi}$\ldots$  & 1  & 87  & 69  & 0.01 \\ 
    \hline
     \multirow{5}{*}{\begin{turn}{90}CUMUL\end{turn}}
       & zehli$\ldots$  & 2  & 15  & 68  & 0.05 \\ 
       & 4ewrw$\ldots$  & 2  & 29  & 68  & 0.04 \\ 
       & harry$\ldots$  & 2  & 29  & 68  & 0.04 \\ 
       & sqtlu$\ldots$  & 2  & 35  & 68  & 0.04 \\ 
       & yiy4k$\ldots$  & 1  & 14  & 69  & 0.02 \\ 
    \hline
     \multirow{5}{*}{\begin{turn}{90}k-FP\end{turn}}
       & \textbf{ykrxn}$\ldots$  & 4  & 62  & 66  & 0.06 \\ 
       & t4is3$\ldots$  & 3  & 42  & 67  & 0.05 \\ 
       & \textbf{wiki5}$\ldots$  & 3  & 55  & 67   & 0.05 \\ 
       & jq77m$\ldots$  & 2  & 54  & 68  & 0.03 \\ 
       & \textbf{newsi}$\ldots$  & 2  & 63  & 68  & 0.03 \\ 
   \end{tabular}}
  \end{threeparttable}
  \vspace{-0.3cm}
\end{table}

We have compared the misclassifications of all three attacks to find sites that
are misclassified by all the classifiers as opposed to sites that at least one
of identified correctly. \autoref{tab:top_sites} shows the top
five \hses ranked by number of misclassifications, where we see a partial overlap of which sites are misclassified the most. This means there is not only variation across websites within a given classifier but also across different classifiers. 

\subsection{Comparison of Website Classification Errors}

\begin{figure}[bth]
 \centering
 \resizebox{0.6\linewidth}{!}{\input{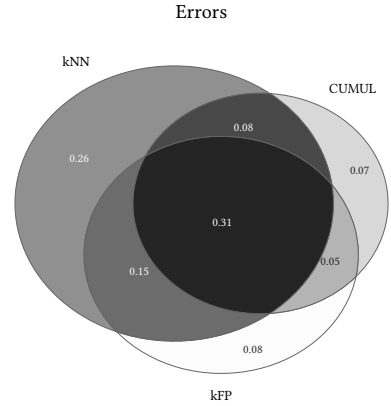}}
 \caption{Scaled Venn diagram of classification errors. Each circle represents the set of prediction errors for a method: \emph{kNN}, \emph{CUMUL} and \emph{kFP}. In the intersections of these circles are the instances that were incorrectly classified by the overlapping methods. 31\% of the erred instances were misclassified by all three methods, suggesting strong correlation in the errors.}
  \label{fig:venn_errors}
\end{figure}

\autoref{fig:venn_errors} shows a scaled Venn diagram of the classification errors. The circles represent the errors made by each of the classifiers, and the intersections represent the fraction of instances misclassified by the overlapping classifiers.
All numbers in the Venn diagram add to one as each number is a fraction of all misclassifications, not a fraction of the misclassifications for a specific classifier. This is to represent how misclassifications are distributed over classifiers and intersections of classifiers. The black region in the center represents the errors that are common to all three classifiers, which accounts for 31\% of all classification errors. This large intersection indicates that classification errors for a given website are correlated and not independent for each classifier. Note that if the errors were independent, the adversary would benefit from employing multiple \wf classifiers; but the correlation suggests that such gains will have limited returns.

\begin{figure}[bth]
 \centering
 \resizebox{0.65\linewidth}{!}{\input{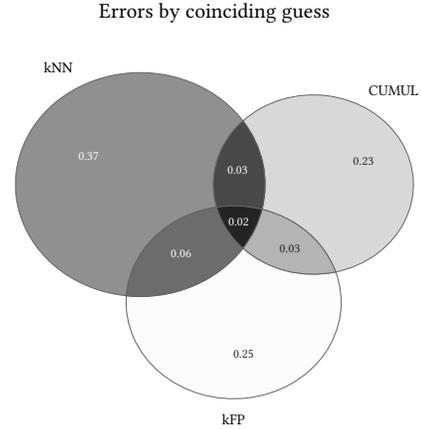}}
 \caption{Scaled Venn diagram of classifications errors by coinciding guess. The
	intersections contain instances that were incorrectly classified with \emph{exactly} the same label by the overlapping classifiers. Only 2\% of the errors were misclassified to the same incorrect site by all three methods, while 85\% were misclassified differently by each method, showing that the methods do err in different ways.}
  \label{fig:venn_coinc_errors}
\end{figure}

The diagram in \autoref{fig:venn_errors} does not take into account whether the classifiers that erred predicted the
same mistaken label or not. In \autoref{fig:venn_coinc_errors}, we depict the
Venn diagram of misclassifications according to the (erroneous) guessed label. The percentage of instances that were mislabeled in the same way by all three classifiers is
substantially smaller: 
only 2\% of the errors are errors that all three classifiers erred with the same predicted label. Interestingly, this small intersection implies that even
though these classifiers err on the same instances
(\autoref{fig:venn_coinc_errors}), they do so in different ways, making different predictions for a given instance.

\subsection{Ensemble Classifier}

In Figure~\ref{fig:venn_errors} 
we observe that more than 25\% of the errors occur in only one of the
methods, and an additional 17\% of errors appear in only two of the
 methods. A third of the errors were misclassified by all three methods.  
 Thus, an \emph{ensemble} classifier that appropriately
combines the three classifiers can achieve higher accuracy than
any individual classifier alone, by correcting  classification errors
that do not occur in all the methods. 

We can estimate the maximum
improvement that such an ensemble could achieve by looking at the potential
improvement of the best classifier. In our case, CUMUL has the greatest accuracy
with 874 errors that could be corrected using kNN or kFP. So if CUMUL did not make these errors, its accuracy would be improved by
$\frac{874}{33,740} = 2.6\%$.  Even though the margin for improvement is small,
we build an ensemble to reduce the dependency of our results on a
single classifier. In addition, by choosing an ensemble we ensure that we are not
underestimating an adversary that combines all the state-of-the-art classifiers. 
We  therefore use the results of the ensemble  to determine fingerprintability, 
and compute  a site's {\bf{\emph{fingerprintability score}} as its F1 score from the ensemble classifier.}

We analyze the overlap in errors and TPs for the three
classifiers for different ensemble methods, as follows:


  {\bf Random}. For each instance,  randomly select one of the
		predictions of the three classifiers. With this method the ensemble
		achieves 79.98\% accuracy.

{\bf Highest confidence}. For each instance, take the prediction of the
		classifier with highest confidence. kFP and CUMUL use Random Forests and SVM respectively, 
        and both output a classification probability for each of the possible classes.
        For kNN we use the distance to the nearest neighbor as the confidence metric.  
        The accuracy was 80.91\% using this method.

{\bf $\bf P_1-P_2$ Diff}. For each instance, use the output of the classifier with the
		greatest difference in confidence between its first and second predictions. We obtained 80.91\% accuracy with this method.

\begin{figure}[b]
 \centering
 \includegraphics[scale=0.5]{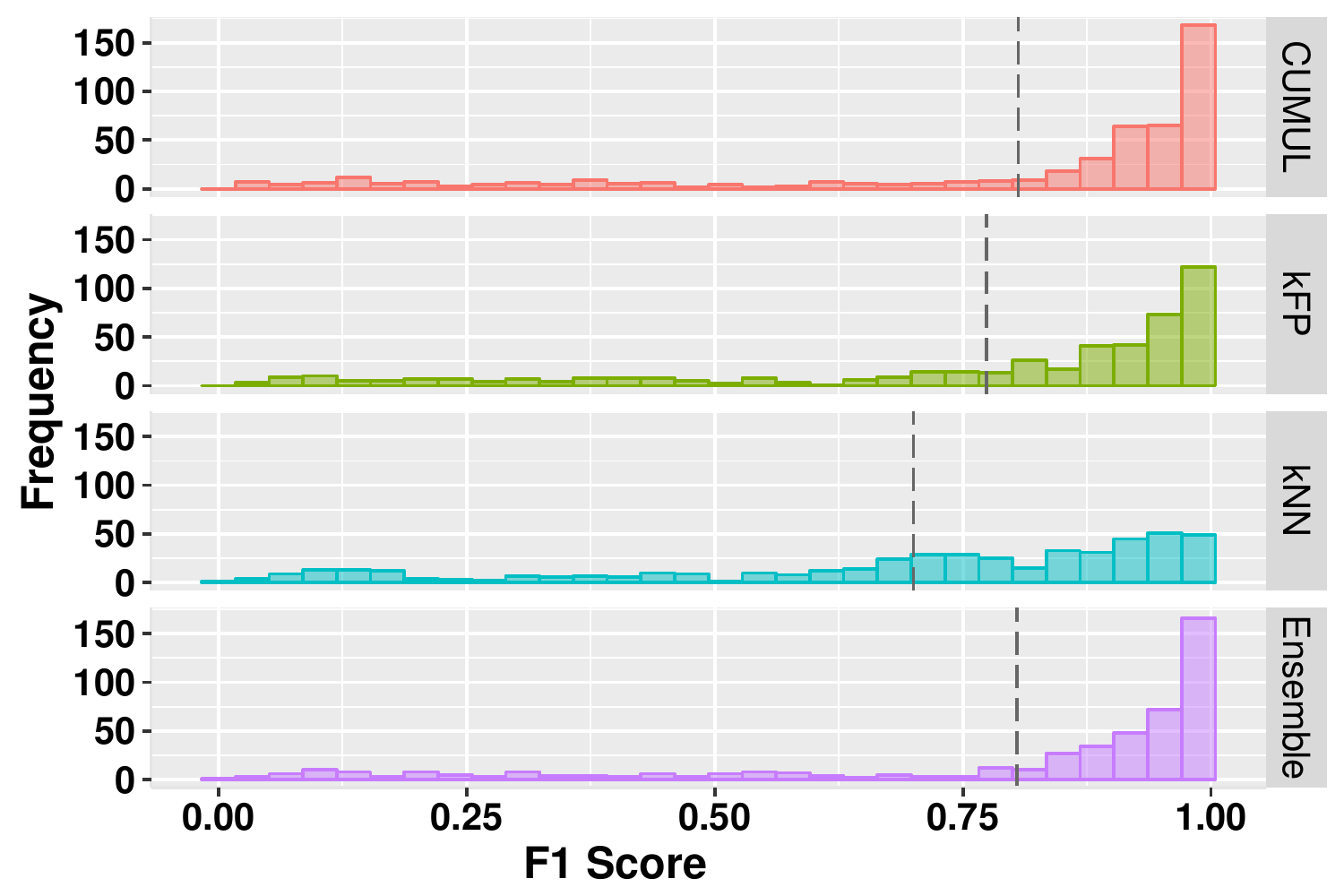}
 \caption{F1 score histograms for each classifier. Vertical dashed lines represent the mean F1 score.}
  \label{fig:fability_histograms_by_classifiers}
\end{figure}

We decided to use the $P_1-P_2$ Diff for the rest of our analysis because it uses most information about the confidence vector. \autoref{fig:fability_histograms_by_classifiers} shows the F1 score histograms for all classifiers including the ensemble. The vertical dashed lines show the mean F1-scores. We note that the ensemble is only marginally better than CUMUL. The main visible difference is in the relative weights of the second and third highest bars: the ensemble improves the F1 score for a subset of instances that in CUMUL contribute to the third bar, and to the second in the ensemble.

In the histograms we can once more see the accuracy variation across sites (horizontally) and across classifiers (vertically). Even though for CUMUL and the ensemble most of the sites have high F1 scores, we see there still are several sites in the low ranges of F1 scores that even CUMUL and ensemble cannot perfectly fingerprint (the ones shown in \autoref{tab:top_sites}). 


\subsection{Sources of Classification Error}


In order to gain insight about the nature of the classifier errors, we performed an exploratory analysis specific to the features of the erred instances. We use  the \emph{total incoming packet size}  as example for illustrating the analysis, because, as we show in the following sections, it is the most salient feature. However, this analysis can as well be applied to any other feature.

\begin{figure}[bt]
 \centering
 \includegraphics[scale=0.25]{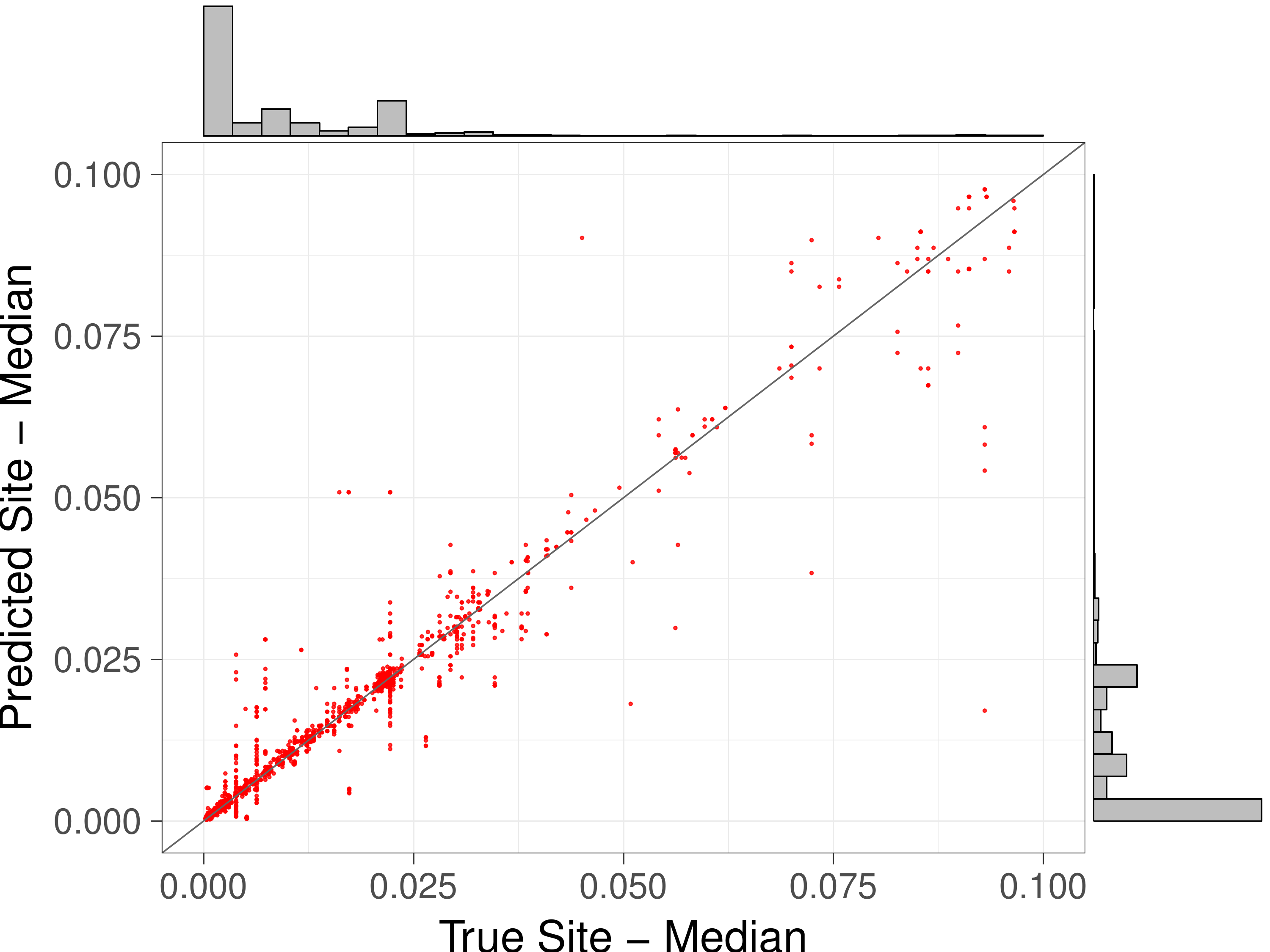}
 \caption{Median of total incoming packet size for misclassified instances (true vs predicted site). We also plot the dashed diagonal line, $y=x$, for comparison. We chose the total incoming packet size for this analysis because it is the most distinguishing feature (see Section~\ref{sec:low-level}).}
  \label{fig:confusion_pairs_median_size}
\end{figure}

In \autoref{fig:confusion_pairs_median_size}, each point represents a misclassified instance, with the $x$ axis value being the median incoming packet size of the `true site' (site the instance truly belongs to), and the $y$ axis value being the median incoming packet size of the `predicted site' (according to the \emph{ensemble} classifier). Note that the total incoming packet sizes have been normalized to the interval $[0,1]$ using Min-Max normalization across all instances. For visualization purposes, we have clipped the range to focus on the region where approximately 80\% of the data points are (101 points were excluded).

\autoref{fig:confusion_pairs_median_size} shows that the median incoming packet sizes of the
predicted and true sites are highly correlated: most of the instances are close
to the diagonal $y=x$ (dashed line), meaning that for most
of the errors, true and predicted sites are similar to each other in terms
of median incoming packet size. In fact, since the median incoming packet size approximates to the median total size of the page, this shows that most of the misclassified pages were confused with pages of similar size. Furthermore, as shown by the histograms most of the misclassifications occur on pages of small sizes, confirming the hypothesis that large pages are easier to identify.

We also measure the deviation of each  instance from its class mean. We use \emph{Z-score}, which indicates the number of standard deviations a sample is away from the mean. The Z-score is a standard statistic that normalizes the deviation from the mean using the class' standard deviation. Unlike the standard deviation, this allows to compare Z-scores between classes with standard deviations that differ by orders of magnitude. This property is suited to our case because the sites in our set have large differences in terms of the total incoming packet sizes.

\begin{figure}[bt]
 \centering
 \includegraphics[scale=0.4]{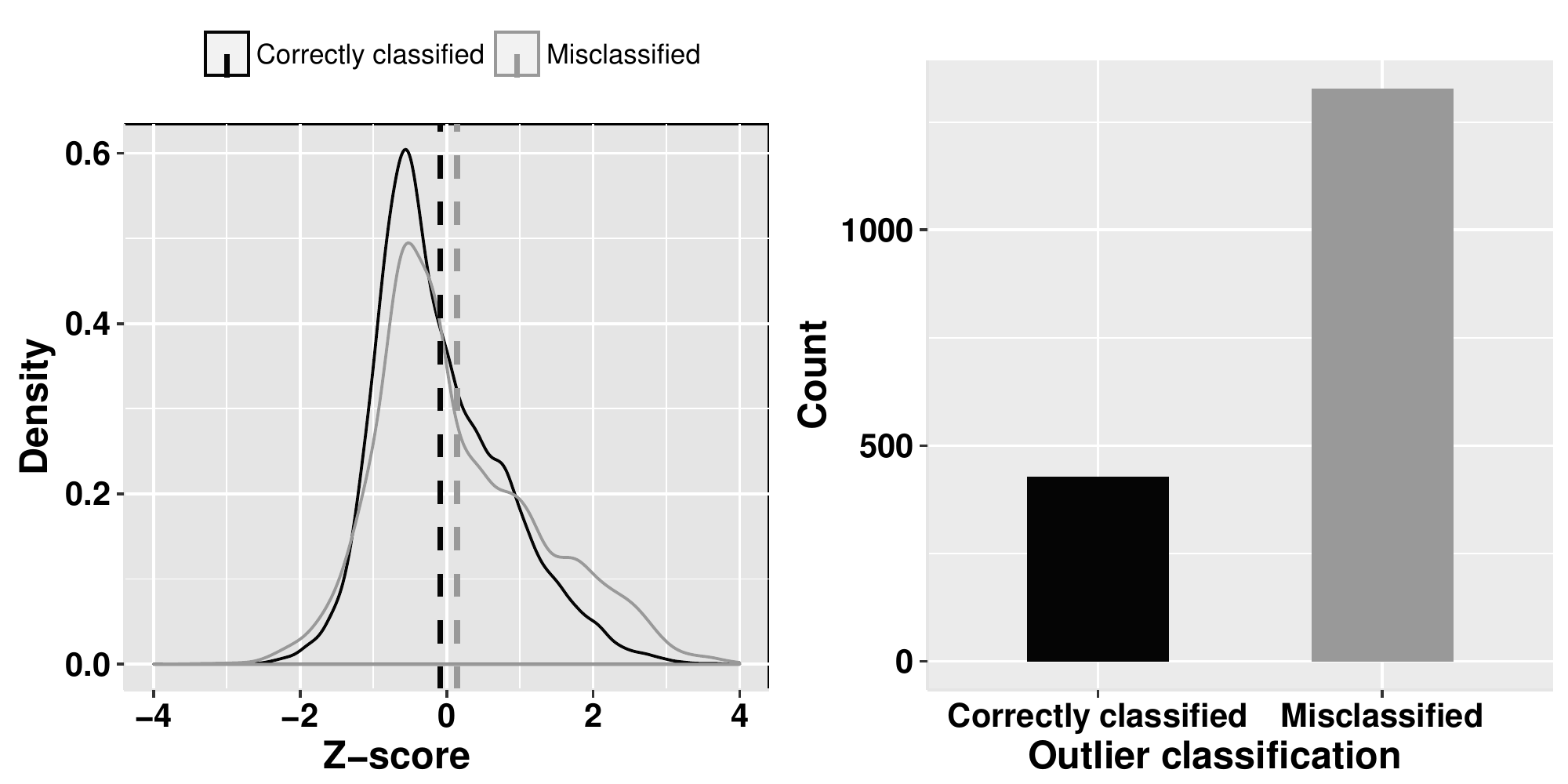}
	\caption{Density plot for absolute value of Z-score distribution of total incoming packet size. Correctly classified (dark gray) and misclassified (light gray) instances are plotted separately to contrast them with respect to their deviation from the class mean.}
 \label{fig:dens_plot_zscore}
\end{figure}

On the left side of \autoref{fig:dens_plot_zscore} we plot the density for the deviation from the median for the total incoming packet size feature. Z-score values around the origin correspond to low-deviation, whereas values far from the origin correspond to high-deviation. We observe that the correctly classified instances are more concentrated in the center, while the misclassified instances are more concentrated in the extremes. This confirms that the instances with higher deviation from their class mean are more likely to be misclassified.

The right subfigure in~\autoref{fig:dens_plot_zscore} shows the number of correctly and erroneously classified  instances for the $1,755$ outliers found in our dataset. We used the Tukey's method for outlier removal based on the inter-quartile range and the first and third quartiles to identify outliers. The bar plot shows that an outlier is three times more likely to be misclassified ($1,327$) than  correctly classified ($428$). An instance is counted as misclassified if it is misclassified by at least one of the classifiers.

~\autoref{fig:dens_plot_zscore} suggests that  variation within a class such as that produced by web page dynamism can be beneficial to induce confusions with other pages.

 
\subsection{Confusion graph}~\label{sec:conf_graph}

Confusion matrices have been used in prior \wf literature to visualize and help understand the nature of confusions~\cite{Hayes2016,Panchenko2016}. However, for a multi-class problem of size 482, the confusion matrix is too large for any visualization to be useful. This can be addressed by using~\emph{confusion graphs} instead, which represent misclassifications as a directed graph~\cite{Yoshida2016}.

To better understand the nature of classification errors we draw a directed graph where nodes represent classes (\hses) and edges represent misclassifications. Source and target nodes of an edge represent true and predicted sites, respectively. The edge weight encodes the misclassification frequency (i.e., number of times the source class is misclassified as the target class).
We have created a confusion graph for CUMUL, which is the best performing classifier in our dataset, 
shown in~\autoref{fig:cumul_confusion_graph} in the Appendix.

The nodes are colored based on the community they belong to, which is determined by the Louvain community detection algorithm~\cite{blondel2008fast}, as implemented in the Gephi graph software. Node size is drawn proportional to the node degree. We observe highly connected communities on the top left, and the right which suggests clusters of \hses  which are commonly confused as each other. Further, we notice several node pairs that
are commonly classified as each other,
forming \emph{ellipses}.

The mean~\emph{outdegree} and ~\emph{indegree} of the graph is 4.9, meaning that, on average, a site is misclassified as 5 distinct sites and confused with 5 distinct sites.
The \hs with the maximum outdegree had 42 outgoing edges, meaning it is misclassified as 42 distinct sites. The \hs with the maximum indegree had 28 incoming edges, meaning it is confused with as many different sites. Interestingly, the same \hs has zero outdegree, i.e., its instances are never misclassified as belonging to another site.

We have looked into the size of the sites for each community in the graph. The sites in the dark green community at the bottom of the graph are all of similar size and significantly larger than all the others, explaining why they are confused between each other and clustered into a community. For the other communities, however, it is not obvious which common features define the community. Further, we discovered that a few of the pairs of sites that form ellipses are false negatives of our duplicates detection in the data cleansing step, while the others require further analysis. We leave a more detailed graph-based analysis of these communities for future work.

We analyze three cases of the symmetry of classifications: 

\begin{itemize}
 \item Symmetrical: Site A is misclassified as other sites and other sites are
misclassified as Site A.
 \item Asymmetrical: One or more sites are misclassified as Site A, but A is
consistently classified as A.
 \item Asymmetrical: Site A is misclassified as one or more other sites, but
other sites are rarely misclassified as A.
\end{itemize}

For each distinct misclassification pair ($A\rightarrow B$) we check whether there is a
symmetric misclassification ($B\rightarrow A$). The total number of misclassifications with
symmetric counterparts:

\begin{itemize}
 \item CUMUL: 74.8\% (4868/6502)
 \item kFP: 73,4\% (5517/7519)
 \item kNN: 80.6\% (8174/10132)
\end{itemize}

The results show the majority of the misclassifications are symmetrical, meaning that there are sets of pages that provide cover for each other, effectively forming anonymity sets .  This suggests that onion services may benefit from designing their site to have features that enable them to join one of those sets. 

\section{Network-Level Feature Analysis} \label{sec:low-level}
We use classifier-independent feature analysis methods to determine which features are better predictors for \wf. Knowing which features are more distinct across classes and less distinct within a class helps us understand which features are important to each \wf method.

\subsection{Methodology}

To analyze the nature of the classification errors we borrow two
concepts from the field of machine learning: \emph{inter-} and
\emph{intra-class} (or \emph{cluster}) variance. In particular, we use
these concepts in the following sense:

The {\bf intra-class variance} of a feature is defined as the variance of its distribution for a certain class, in this case a site. It quantifies how much the feature varies among instances of the class. In \wf, low intra-class variance indicates a feature remains stable across different visits to the same page.
  
{\bf Inter-class variance} is a measure of how much a feature varies across different classes. We define it as the variance of the averages of the feature for each class. That is, we create a vector where each coordinate aggregates the instances of visits to a site by averaging their feature values. Then, we calculate the inter-class variance as the variance of that vector. In \wf, high-inter-class variance means that websites are very distinct from each other with respect to that feature.


In Section~\ref{sec:misclassif} we have shown evidence that both inter- and intra-class variance play a role as the cause of classification errors:  misclassified pages have similar sizes to the pages they are confused with, and slightly larger variance in size than correctly classified ones. To rank features by taking into account both intra- and inter-class variance, we use the relative difference between the inter- and intra-class variance, where we define relative difference as: $d(x,y) = (x-y)/((x+y)/2)$. This formula normalizes the differences by their mean to values between 0 and 2, where features with a relative difference close to 0 are similar and features with a relative difference close to 2 are far apart. This allows features of different scales to be compared. We consider features that are close to 2 better predictors, as they have a relatively higher inter-class variance than intra-class variance. 

Many of the features that appear as most predictive for the considered classifiers are directly related to the size of a site (e.g., the number of packets). Further, the misclassifications described in Section \ref{sec:misclassif} show that the smaller sites are more likely to be misclassified. In addition to running feature analysis on the entire dataset, we also look only at the small sites to determine which other features have predictive value.

We start with an analysis of the network-level features used by the three fingerprinting attacks detailed in Section \ref{sec:related} and analyzed in Section \ref{sec:misclassif}. 
Most traditional applications of feature analysis aim to reduce the dimensionality of the data to more efficiently classify instances. Instead, the goal of our feature analysis is to determine which features can be modified to trick a classifier into misclassifying an instance. Unlike many adversarial machine learning problems with the same goal, this analysis lacks knowledge of the specific classifier (or even the classification algorithm) used for fingerprinting, as there are many different classifiers in the literature to consider, and the site should ideally be hard to classify for all of them. In addition to the wide variety of classification techniques available in the current literature,  novel classification techniques could be easily developed by an adversary. 

\begin{figure}[h]
\includegraphics[width=\columnwidth]{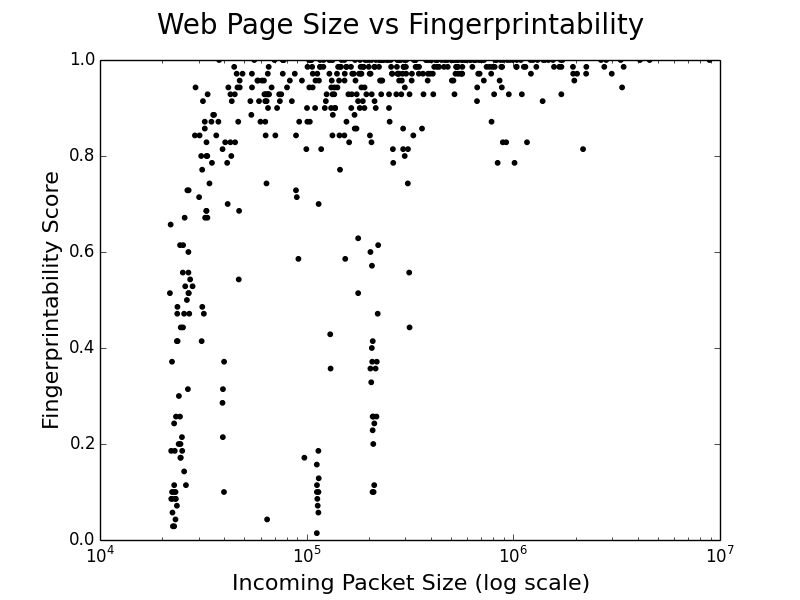}
\caption{Larger sites are easily fingerprinted while results are mixed for smaller sites. Note also the vertical clusters of sites with low fingerprintability that are similar in size. Incoming packet size (in bytes) is plotted in log scale.} 
\label{fig:size_vs_fability}
\end{figure}

Therefore, the network-level feature analysis we present here is classifier-independent. That is, we use only information about the feature values themselves and do not use classification methods to determine the importance of the features. Figure \ref{fig:size_vs_fability} shows the relationship between how likely a site is to be fingerprinted vs its size. All of the larger sites have high fingerprintability scores, while the scores of smaller sites are much more varied. 

In a \wf attack, only features based on the traffic traces are available to the adversary. Each attack uses a distinct set of features derived from these traces and as a result the exact feature analysis varies.

This analysis is classifier independent, meaning no classification techniques were performed on the dataset prior to this analysis and the results do not rely on any specific classification algorithm or task. We cannot, however, perform any feature analysis that is completely independent from the \wf methods, as the types of features we analyze rely on the features chosen by each method. For each attack, however, we can determine which features are most predictive. 

\subsection{Network-Level Feature Results}

Here we analyze which network-level features are the best predictors in state-of-the-art \wf attacks. 

\subsubsection{CUMUL}

The first group of features we consider come from the CUMUL attack. There are two types of features used in CUMUL: direct size features (Table \ref{tab:cumullow}) and interpolated features. The interpolated features are  formed by the number of bytes and packets in each direction and 100 interpolation points of the cumulative sum of packet lengths (with direction). 
We calculate the inter and intra-class variance for each of these features. The direct size features are the most important to classification (Table \ref{tab:cumullow}). We found that the interpolated features are more predictive at the end of the trace than the beginning, with the minimum relative difference (0.37) being from the very first interpolated feature and then increasing to the greatest relative difference (1.51) being the last interpolated feature from the very end of the trace. 


\begin{table}[h]
\begin{tabular}{l r}
{\bf Feature Name} & {\bf Relative Diff} \\ \hline
Total Size of all Outgoing Packets & 1.605 \\
Total Size of Incoming Packets & 1.520 \\
Number of Incoming Packets & 1.525 \\
Number of Outgoing Packets & 1.500
\end{tabular}
\caption{Network-Level Feature Variance Analysis for CUMUL Method. These features had a higher relative difference than most of the interpolated features and alone are great predictors.}
\label{tab:cumullow}
\end{table}

\subsubsection{k-fingerprinting} 

The next group of features we look at come from the k-fingerprinting attack. The features used in the k-fingerprinting attack are more varied as well as more straightforward than those in CUMUL. They include not only features that give information about the size and number of packets, but also the timing of the packets. The features with the highest inter-class to intra-class variance ratio are shown in Table \ref{tab:kfplow}.

The feature analysis we present here is similar to the original analysis presented with the method by the authors, but without the use of any classification technique. Further, we also look at which features are more predictive for small sites, as we see that misclassifications are much more common for smaller sites. 

Table \ref{tab:kfplow} shows that features correlated to the total size of a site (e.g. \# of outgoing packets) have the highest relative difference and thus are among the top features. This result is consistent with the analysis done by Hayes and Danezis\cite{Hayes2016} on the same set of features. 

When only smaller sites are analyzed however, standard deviation features become important. In Section~\ref{sec:misclassif}, we show that large sites are easily identified, and the fact that size features are very predictive is not at all unexpected. However, that standard deviation features are top features for the smaller sites implies that the dynamism of the site makes a difference, as small dynamic sites are generally the least fingerprintable.  

\begin{table}[h!]
\begin{tabular}{p{140pt} r}
{\bf Feature name }& {\bf Relative Diff} \\ \hline
{\bf All Sites} &\\
Percent incoming vs outgoing & 1.895 \\ 
Average concentration of packets & 1.775 \\ 
\# of outgoing packets	& 1.740 \\ 
Sum of concentration of packets	& 1.740 \\ 
Average order in	& 1.720 \\ 
{\bf Smallest 10\% of Sites}&\\
Percent incoming vs outgoing & 1.951 \\
Average concentration of packets & 1.944 \\
Standard deviation of order in & 1.934 \\ 
\# of packets	& 1.927 \\ 
\# of packets per second	& 1.927 \\

\vspace{5pt}

\end{tabular}
\caption{Network-level feature analysis for kFP method.}
\label{tab:kfplow}
\end{table}


\subsubsection{kNN}

The last set of features are those of the kNN attack. Like with the other classifiers, we find that the most important features are those that relate to the size of the traffic flow. In this case, we find that almost all of the top predictive features (with the highest relative difference) are related to ``packet ordering'' -- which in practice acts as proxy for the size of the flow. 

The packet ordering feature is computed as follows: for each outgoing packet $o_i$, feature $f_i$ is the total count of all packets sent or received before it. Essentially, these features measure the ordering of incoming and outgoing packets.Note that not all sites, however, have the same number of outgoing packets. Therefore if the end of the number of outgoing packets is less than some $n$ (we use $n=500$ to be consistent with the original implementation), the rest of the features are filled in with zero or null values. Similarly, some sites may have over $n$ outgoing packets. If this is the case, the packets over the $n^{th}$ packet are ignored. Similar to the features used in CUMUL, we observed that the later features in this sequence are more important, this is because for most sites (size $< n$) they are zero and thus these features are a proxy for the total size of the site. 

The only other feature-type with high relative difference between inter and intra-class variance is the number of packets (1.96), a  direct measure of the size of the site.






\section{Site-Level Feature Analysis} \label{sec:high-level}

In \wf attacks, the adversary records the network traffic between a user and Tor, and analyzes its features  to identify the site that was visited. Network-level features and their relative contribution to  fingerprintability are, however, not informative for \hs designers who may want to craft their site to be robust against \wf attacks. 
To gain insight into which design choices make sites vulnerable to attacks, and how websites can be designed with increased security, we need to look at the features at a site-level. 

In this section we investigate which site-level features correlate with more and less fingerprintable sites. Site-level features are those that can be extracted from a web page itself, not from the traffic trace. Driven by adversarial learning, we investigate the task of causing misclassifications for any set of network-level features and any classification method. This information can help sites design their web pages for low fingerprintability, and also assist in developing more effective server-side defenses.

\subsection{Methodology }
Site-level features are extracted and stored by our data collection framework as explained in Section~\ref{sec:data_collection}. The list of all site-level features considered can be found in~\autoref{tab:hifeats} (in the Appendix). 

We build a random forest regressor that classifies easy- and hard-to-fingerprint sites, using the {\bf{fingerprintability scores}} (the F1 scores from the ensemble classifier described in Section \ref{sec:misclassif}) as labels, and considering site-level features.  
We then use the fingerprintability regressor as a means to determine which site-level features better predict fingerprintability.


In this section we aim to understand which site-level features are more prevalent in the most and least fingerprintable sites. For the sake of this feature analysis, we remove the middle tier of sites, defined as those with a fingerprintability score in (0.33, 0.66). 44 sites in our dataset were assigned a mid-ranged F1-score, leaving 438 sites for this analysis. 


The next challenge is that the high and low-fingerprintability classes are unbalanced, because of the disproportionately higher number of easily identifiable sites compared to the amount of sites that are hard to identify. Recall that a full 47\% of sites in our dataset have a fingerprintability score greater than 95\%. A regressor trained with such unbalanced priors will be biased to always output a prediction for of ``very fingerprintable,'' or values close to 1, and therefore any analysis on the results would be meaningless. To perform the feature analysis, we remove randomly selected instances from the set of more fingerprintable sites, so that it is balanced in size with that of low fingerprintability. 

We train a random forest regressor using the features from Table~\ref{tab:hifeats}. 
We use the feature weights from the regression to determine which of these site-level features are most predictive of sites that are easily fingerprinted.
We use the information gain from the random forest regression to rank the importance of the site-level features in making websites more or less fingerprintable. 

While in its current state this regression is only useful for feature analysis, this could be extended into a tool that allows sites to compute their fingerprintability score, and be able to determine if further action is needed to protect their users from \wf attacks. 

\subsection{Results}

\begin{figure}[h]
\includegraphics[width=\columnwidth]{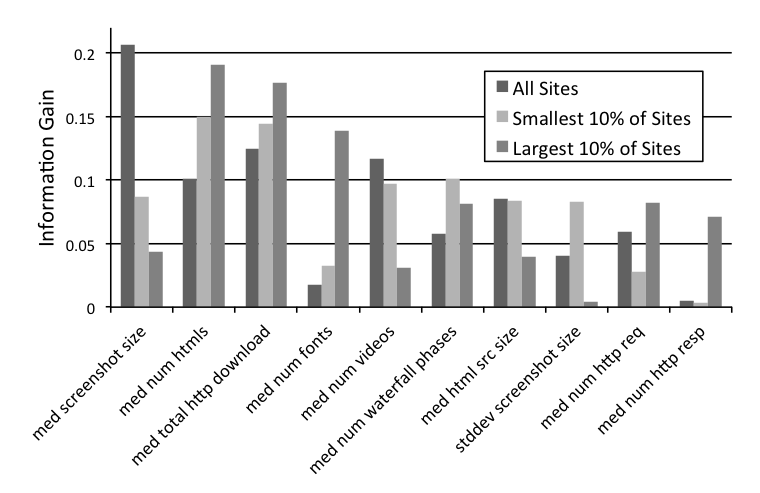}
\caption{Most important features by information gain. Features related to the size of a site are important.}
\label{fig:sitelevel}
\end{figure}

Figure~\ref{fig:sitelevel} shows the results of the analysis. We see that  features associated with the size of the site give the highest information gain for determining fingerprintability when all the sites are considered. Among the smallest sites, which are generally less identifiable, we see that standard deviation features are also important, implying that sites that are more dynamic are harder to fingerprint.

Additionally, Table~\ref{tab:high_level_summary} shows how different the easy- and hard-to-fingerprint sets of sites are in terms of total HTTP download size, a straightforward metric for the size of a site. The median site size for the 50 most fingerprintable sites is almost 150 times larger than the median size of the harder to classify sites. The standard deviation of the total site size for the most and least fingerprintable sites, relative to their size, is similarly distinct, showing the most fingerprintable sites are less dynamic than the 50 least fingerprintable sites. That is, they are less likely to change between each visit. 

\begin{table}[h]
\begin{tabular}{rcc}
Total HTTP Download Size &50 Most & 50 Least \\ \hline
Median Std Dev & 0.00062 & 0.04451\\
(normalized by total size) & &  \\
Median Size & 438110 & 2985
\end{tabular}
\caption{Differences in the most and least fingerprintable sites. The 50 most fingerprintable sites are larger and less dynamic than the 50 least fingerprintable sites.}
\label{tab:high_level_summary}
\end{table}

\begin{figure}[h]
\includegraphics[width=\columnwidth]{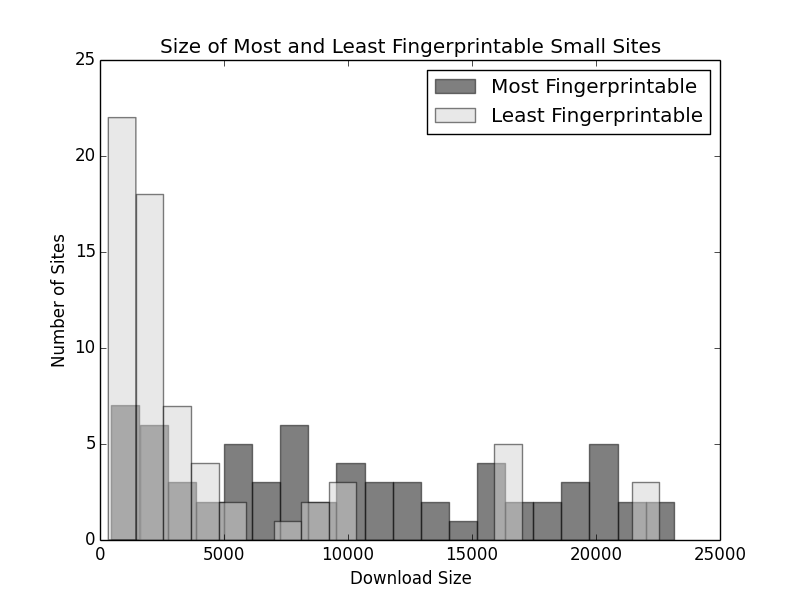}
\caption{Distribution of  sizes for the most and least fingerprintable sites, considering only the sites smaller than 25,000 bytes.}
\label{fig:hist}
\end{figure}

While the smallest sites are less fingerprintable, some are still easily identified. Figure \ref{fig:hist}  shows the distribution of sizes considering only the smallest sites, distinguished by whether they have a high or low fingerprintability score. We can see that the least fingerprintable sites are clustered in fewer size values, while the most fingerprintable are more spread, meaning that there are fewer sites of the same size that they can be confused with.

\pagebreak

\section{Implications for \hs design}~\label{sec:discussion}

Overall, our analysis showed that most \hses are highly vulnerable to \wf attacks. Additionally, we found that larger sites are more susceptible to \wf attacks. Larger sites were more likely to be perfectly classified by all attacks while many smaller sites were able to evade the same attacks by inducing misclassifications. 

We also observed that the small sites that are harder to identify also have a high standard deviations for many site-level and network-level features, implying that dynamism plays a role in why these sites are less identifiable. While our results show that small size is necessary, it is not sufficient. As a result, our recommendation for \hs designers is ``make it small and dynamic.''

Most \wf defenses rely on some form of padding, that is, adding spurious traffic and therefore increasing the download size. Our analysis, however, shows that this type of defense may not be robust when features such as download size become sparse. Often, these defenses are tested against a single attack with a single feature set and a specific classification algorithm. We see, though, that classification errors do not always coincide for different attacks, and argue that any \wf defense needs to be tested against a range of state-of-the-art attacks, preferably relying on different algorithms and feature sets, in order to provide more general guarantees of its effectiveness.

As a case study, we consider the results that our ensemble classifier achieved in identifying SecureDrop sites. These sites are \hses that are running the SecureDrop software, a whistleblower submission system that allows journalists and media publishers to protect the identities of their sources. Given the sensitive nature of the service that they provide and the nation-state adversaries that they may realistically face, these SecureDrop sites have strong anonymity requirements.  

Our dataset contained a SecureDrop site owned by `Project On Gov't Oversight' (POGO)\footnote{\url{https://securedrop.pogo.org}}. The SecureDrop site had an F1-Score of 99\%, meaning that it is much more vulnerable to \wf attacks than the average \hs site. 

There were other SecureDrop sites present in our initial dataset, associated with The New Yorker, The Intercept and ExposeFacts. These sites were flagged as duplicates of the POGO SecureDrop site and thus removed during the data processing stage. Since they were identified as duplicates, all these SecureDrop sites have very similar characteristics and can thus be expected to be identifiable at a similarly high rates as the POGO site. In particular, we noted that these pages embed images and use scripts and CSS styles that make them large and therefore distinguishable. 

It can be argued that the existence of various similar SecureDrop sites creates an anonymity set and makes some sites cover up for each other. On the other hand however, it may be enough for the adversary to ascertain that the user is visiting \textit{a} SecureDrop site for the anonymity of the source to be compromised. 

We did a small, manual analysis of some of the most and least fingerprintable sites (by F1 score) to see if there were any strong correlations with content. We found that pages at the bottom end of the spectrum were smaller and simpler (a hidden wiki, a listing of a directory, nginx config page, etc.) whereas the most fingerprintable pages were larger and more complex (a bitcoin faucet site, a forum, the weasyl art gallery site, propublica, a Russian escort service site). Pages in the middle of the spectrum varied, but were often login pages. It is worth pointing out that the onion services ecosystem has a 90's, GeoCities ``look,'' where pages tend to be simple HTML and sites that do not follow this aesthetic will stand out. 







\section{Limitations and Future Work}~\label{sec:limitations_future}



With 482 onion sites, this is the largest \wf study of \hs sites.
Even so, our results may not be representative of the entire \hs universe. We made our best effort to collect as many \hs URLs as possible using ~\texttt{ahmia.fi}. While there are more effective methods to collect .onion addresses, such as setting up a snooping Hidden Service Directory~\cite{sanatinia2016honions}, they are ethically questionable.

Our data is a snapshot of the \hses space over 14 days. As the \hses change constantly,  and fingerprintability depends not just on individual sites but the whole set, the dataset and the analysis should be updated regularly for a diagnosis of current levels of fingerprintability.

As new \wf attacks are proposed, features that are important to fingerprintability now may become less so, especially if defenses are introduced or if the design of websites changes.
The methods introduced in this paper for extracting features and understanding what makes certain sites identifiable, however, are a lasting and relevant contribution. In particular, we argue that the effectiveness of a proposed defense should be examined not only on average, but that it should account for possible disparate impact on different sites depending on their features. For example, even if a defense significantly lowers the average accuracy of a \wf attack, it could be that certain sites are always correctly identified, and thus left unprotected by the defense.
%
We also point out that we focus on whether a site blends well with other sites, triggering frequent misclassifications in the context of \wf attacks, and that the effectiveness of using such techniques as basis for defending against \wf, has dependencies on the actions taken by other \hses. 




Our data collection methodology follows standard experimental practices in the \wf literature when crawling only home pages. On the one hand, limiting the evaluation to home pages (rather than including all inner pages of a site) reduces the classification space and gives an advantage to the adversary compared to considering that users may directly browse to the inner pages of a site. We argue that a fraction of users will still first land on the homepage of a site before visiting inner pages and thus this adversarial advantage is not unrealistic. We also note that the link structure of inner pages in a website can be exploited to improve the accuracy of \wf attacks. 

Compared to using~\texttt{wget},~\texttt{curl} or headless browsers, our Tor Browser based crawler better impersonates a real browser, limiting the risk of differential treatment by \hses. Still, it is possible detect the presence of Selenium based automation using JavaScript.

The adversary can sanitize training data by taking measures such as removing outliers, but cannot do so for test data. Since we measure an upper bound for the fingerprintability of websites, we sanitize the whole dataset including the test data. Note that this is in line with the methodology employed in prior work~\cite{Wang2013, Panchenko2016}.

We acknowledge that redesigning a site to be small and dynamic, as suggested best practice by our analysis, may not be an option for some sites for a variety of reasons. This is a limitation of our approach to countermeasures, but might be a limitation to \wf defenses in general, as large sites are easily identified by \wf attacks.
However, we believe that our results can inform the design of application-layer defenses that alter websites in order to perturb site-level features~\cite{Cherubin2017}. This would allow to optimize existing application-layer defenses by focusing on the features that our site-level feature analysis has identified as most identifying, thus reducing the performance that these defenses incur in Tor.

Previous studies on \wf have shown that data collected from regular sites get stale over time, namely, the accuracy of the attack drops if the classifier is trained on outdated data~\cite{Juarez2014}. For \hses, Kwon et al.\ did a similar experiment and showed that \hses change at a lower rate than regular sites and do not get stale as quick~\cite{Kwon2015}. For this reason, in this paper, we assume the adversary can keep an updated database of \wf templates.

Reducing the accuracy of \wf attacks can be framed as an adversarial learning problem. A webpage can be redesigned to modify its site-level features (especially those that contribute the most to fingerprintability) to trick the classifier into making a misclassification. In future work we plan to tackle finding efficient ways to altering these website features to launch~\emph{poisoning attacks} against \wf classifiers~\cite{huang2011adversarial} under constraints such as bandwidth, latency and availability. 

Finally, we acknowledge that the random forest regression method to determine the fingerprintability of a webpage given only web-level features is currently useful only for feature analysis. This is due to a number of factors, such as removing the middle of the spectrum sites and balancing the priors. Although there are a few challenges and limitations, creating an accurate tool that can determine if a site will be easily fingerprinted from only  site-level features would be very valuable to \hses.

\section{Conclusion}~\label{sec:conclusion}

Our work intends to change the way that we build and analyze \wf attacks and defenses, and differs from previous \wf contributions in several ways. We do not propose a new attack algorithm (with the exception, perhaps, of the ensemble method) or an explicit defense, but study instead what makes certain sites more or less vulnerable to the attack. We examine which types of features, with intentional generality, are common in sites vulnerable to \wf attacks. 

 This type of analysis is valuable for \hs operators and for designers of \wf defenses. A \wf countermeasure may have a very disparate impact on different sites, which is not apparent if only average accuracies are taken into consideration. Further, we note that from the perspective of an \hs provider, overall accuracies do not matter, only whether a particular defense will protect their site and their users. Our results can guide the designers and operators of \hses as to how to make their own sites less easily fingerprintable, in particular considering the results of the feature analyses and misclassifications. For example, we show that the larger sites are reliably more identifiable, while the hardest to identify tend to be small and dynamic. 

This work is also a contribution to adversarial machine learning. Most work in adversarial learning focuses on attacking a specific algorithm and feature set, but in many privacy problems this model does not fit. Our study investigates methods to force the misclassification of an instance regardless of the learning method.

\begin{acks}

This work was funded in part by the National Science Foundation (1253418) and a senior postdoctoral fellowship from KU Leuven (SF/15/007). In addition, this work was supported by the European Commission through KU Leuven BOF OT/13/070, H2020-DS-2014-653497 PANORAMIX and H2020-ICT-2014-644371 WITDOM. Marc Juarez is funded by a PhD fellowship of the Fund for Scientific Research - Flanders (FWO).

\end{acks}

\bibliographystyle{ACM-Reference-Format}
\balance
\bibliography{references}

\appendix
\small

\onecolumn

\section{Site level features}\label{app:a}

\autoref{tab:hifeats} shows the site-level features and statistic used to aggregate each site-level features within a site class. We followed the feature extraction step outlined in Section~\ref{sec:data_collection} to obtain the site-level features. Here we present a more detailed overview of feature extraction for different site-level feature families.

\newcommand{\specialcell}[2][c]{%
  \begin{tabular}[#1]{@{}c@{}}#2\end{tabular}}

\begin{table}[!b]
\caption{Site-level features and statistics used to aggregate them across download instances. Nominal and binary features such as \texttt{Made with Wordpress} are aggregated by taking the most frequent value (i.e. mode) of the instances. Quantitative features such as \texttt{Page load time} are aggregated using median, as is is less sensitive to outliers than the statistical mean.}
\small
\vspace{0.5cm}
\begin{tabular}{p{140pt}|c|c|l}

{\bf Feature} & {\bf Median} & {\bf Mode} &  {\bf Description}  \\ \hline
\texttt{Number of HTTP requests} & \reddot & & Number of HTTP requests stored by the browser add-on \\
\texttt{Number of HTTP responses} &  \reddot & & Number of HTTP responses stored by the browser add-on \\
\texttt{Has advertisement} &  & \reddot &  HTTP request matching EasyList~\tablefootnote{~\url{https://easylist.to/easylist/easylist.txt}} \\
\texttt{Has tracking/analytics} &  & \reddot & HTTP request matching EasyPrivacy~\tablefootnote{~\url{https://easylist.to/easylist/easyprivacy.txt}}\\
\texttt{HTML source size} & \reddot &  &  Size (in bytes) of the page source \\
\texttt{Page load time} & \reddot &  &  As determined by Selenium \\
\texttt{Made with Django} &  & \reddot & As determined by \texttt{generator} HTML meta tag  \\
\texttt{Made with Dokuwiki} &  & \reddot & As determined by \texttt{generator} HTML meta tag \\
\texttt{Made with Drupal} &  & \reddot & As determined by \texttt{generator} HTML meta tag \\
\texttt{Made with Joomla} &  & \reddot & As determined by \texttt{generator} HTML meta tag \\
\texttt{Made with MediaWiki} &  & \reddot & As determined by \texttt{generator} HTML meta tag \\
\texttt{Made with OnionMail} &  & \reddot & As determined by \texttt{generator} HTML meta tag \\
\texttt{Made with phpSQLiteCMS} &  & \reddot & As determined by \texttt{generator} HTML meta tag \\
\texttt{Made with vBulletin} &  & \reddot & As determined by \texttt{generator} HTML meta tag \\
\texttt{Made with WooCommerce} &  & \reddot & As determined by \texttt{generator} HTML meta tag \\
\texttt{Made with Wordpress} &  & \reddot & As determined by \texttt{generator} HTML meta tag \\
\texttt{Made with CMS} &  & \reddot & True if any of the ``Made with...'' features above is true \\
\texttt{Number of audio} &  \reddot & & As determined by the \texttt{Content-Type} HTTP response header \\
\texttt{Number of domains} &  \reddot & & As determined by the \texttt{Content-Type} HTTP response header \\
\texttt{Number of redirections} &  \reddot & &  As determined by the presence of \texttt{Location} HTTP response header \\
\texttt{Number of empty content} & \reddot & & Number of HTTP responses with \texttt{Content-Length} equal to zero \\
\texttt{Number of fonts} &  \reddot & & As determined by the \texttt{Content-Type} HTTP response header \\
\texttt{Number of HTML resources} &  \reddot  & & As determined by the \texttt{Content-Type} HTTP response header \\
\texttt{Number of images} &  \reddot  & & As determined by the \texttt{Content-Type} HTTP response header \\
\texttt{Number of other content} &  \reddot & & As determined by the \texttt{Content-Type} HTTP response header \\
\texttt{Number of scripts} &  \reddot & & As determined by the \texttt{Content-Type} HTTP response header \\
\texttt{Number of stylesheets} &  \reddot & & As determined by the \texttt{Content-Type} HTTP response header \\
\texttt{Number of videos} &  \reddot & & As determined by the \texttt{Content-Type} HTTP response header \\
\texttt{Number of waterfall phases} &  \reddot & & Approximate number of HTTP waterfall chart phases as determined \\ & & & by switches from request to response or response to request.  \\
\texttt{Screenshot size} & \reddot &  & Size (in bytes) of the screenshot saved by Selenium \\
\texttt{Page weight} & \reddot &  & Sum of the HTTP response sizes (in bytes) \\
\texttt{Total request size} & \reddot &  & Sum of the HTTP request sizes (in bytes) \\
\end{tabular}
\label{tab:hifeats}
\end{table}

\newpage
\section{Confusion Graph for CUMUL}\label{app:b}

\begin{figure}[h!]
\centering
\vspace{1cm}
\includegraphics[width=\columnwidth,height=0.77\textheight]{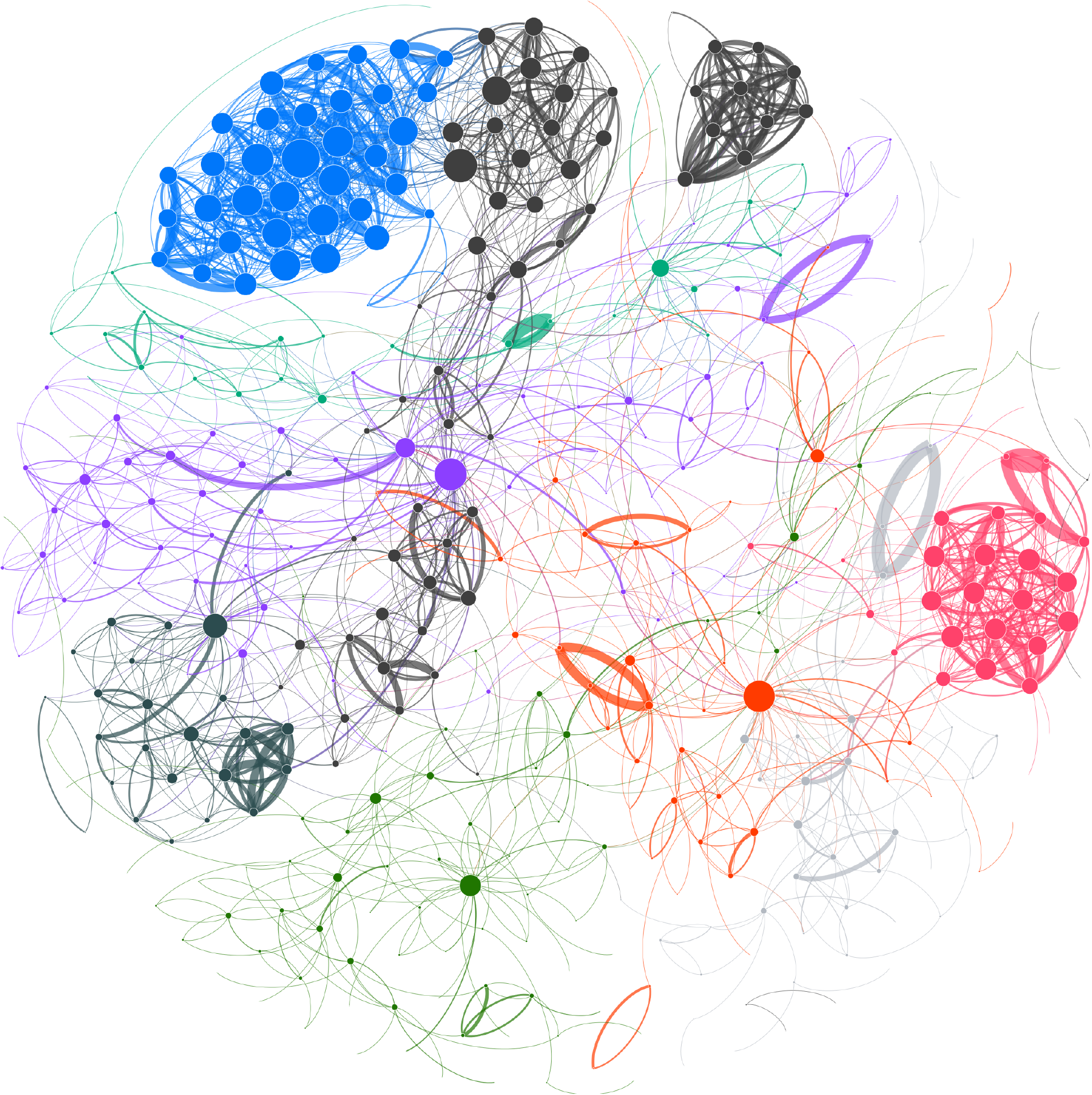}

\caption{Confusion graph for the CUMUL classifier drawn by Gephi software using the methodology explained in Section~\ref{sec:conf_graph}. Nodes are colored based on the community they belong to, which is determined by the Louvain community detection algorithm~\cite{blondel2008fast}. Node size is drawn proportional to the node degree, that is, bigger node means lower classification accuracy. We observe highly connected communities on the top left, and the right which suggests clusters of \hses  which are commonly confused as each other. Further, we notice several node pairs that
are commonly classified as each other,
forming ellipses.} 
\label{fig:cumul_confusion_graph}
\end{figure}

\end{document}